# Numerical Study of Free-Surface Electrohydrodynamic Wave Turbulence

Igor A. Dmitriev, Evgeny A. Kochurin, and Nikolay M. Zubarev

*Abstract*—Direct numerical simulation of three-dimensional chaotic motion of a dielectric liquid with a free surface under the action of external horizontal electric field is carried out. The numerical model takes into account the effects of surface tension, viscosity, and external isotropic random forcing acting at large scales. A transition from dispersive capillary wave turbulence to quasi-isotropic non-dispersive EHD surface turbulence with increase of the external electric field strength is numerically observed for the first time. At the regime of developed EHD wave turbulence, the total electrical energy is found to be much greater than the energy of capillary waves, i.e., electrohydrodynamic effects play a dominant role. At the same time, anisotropic effects are detected that lead to the generation of capillary wave packets traveling perpendicular to the external field direction. Despite the revealed anisotropy, the calculated spectrum of EHD wave turbulence is in very good agreement with the analytical spectrum obtained on the basis of dimensional analysis of weak turbulence spectra.

## I. Introduction

IT is known that nonlinear wave systems can pass into a regime of stationary chaotic motion (wave turbulence) in the result of resonant wave interactions [1]. A theory studying statistically such wave systems is the weak turbulence theory developed by Zakharov and co-authors [1,2]. An important achievement of the weak turbulence theory is the exact analytical solution of kinetic equations describing the nonlinear interaction of waves. The solutions known as the Kolmogorov-Zakharov (KZ) spectra describe the stationary transfer of energy into small or large scales (direct or inverse energy cascades, respectively). To date, the most studied type of wave turbulence is the turbulence of capillary waves propagating along the free surface of a liquid, which is first theoretically investigated in the work of Zakharov and Filonenko [2]. The KZ spectrum for dispersive capillary wave turbulence (also known as the Zakharov-Filonenko spectrum) is usually written in terms of the surface elevation spectrum: $S(k) = 2\pi k |\eta_\mathbf{k}|^2$, with $\eta_\mathbf{k}$ being the Fourier image of surface profile $\eta(x, y)$. The KZ spectrum for capillary waves reads [2]

$$S(k) = C_{KZ} \cdot P^{1/2} (\sigma/\rho)^{-3/4} k^{-15/4}, \quad k = |\mathbf{k}|, \quad (1)$$

where $\mathbf{k} = \{k_x, k_y\}$ is the wave vector, $C_{KZ}$ is the KZ constant, $P$ is the rate of energy dissipation per unit surface area (energy flux), $\sigma$ and $\rho$ are the surface tension and density of the fluid, respectively. The power dependence on $P$ in the spectrum (1) with the exponent 1/2 reflects the resonant character of three-wave interactions. In terms of spectral energy density $E(k)$, the spectrum (1) is rewritten as

$$E(k) = (\sigma/\rho) k^2 S(k) = C_{KZ} \cdot P^{1/2} (\sigma/\rho)^{1/4} k^{-7/4}.$$

To date, the Zakharov-Filonenko spectrum (1) for capillary surface waves has been confirmed with high accuracy both experimentally [3,4] and numerically [5-7].

Currently, the least studied type of surface wave turbulence is the non-dispersive electro- or magnetohydrodynamic (EHD or MHD, respectively) turbulence arising under the action of an external electric (or magnetic) field directed tangentially to the unperturbed boundary of a liquid. Until recently, studies of the nonlinear dynamics of liquid boundaries in electric fields were restricted to consideration of only coherent structures, such as solitary waves or wave collapses [8-13]. The chaotic dynamics of free surface of a magnetic fluid under the action of external magnetic field (MHD surface turbulence) was first studied experimentally in [14,15], where it was shown that the surface turbulence spectrum deviates from spectrum (1) with increasing external magnetic field. A complete theoretical explanation of this phenomenon has not been proposed yet. Previously, EHD and MHD wave turbulence was numerically studied only in plane-symmetric and anisotropic regimes of motion [16-18]. To date, it has not been clarified whether EHD or MHD wave turbulence can be realized in the isotropic case.

The dimensional analysis of weak turbulence spectra [19] allows obtaining estimation for the turbulence spectrum of non-dispersive EHD wave turbulence developing at free surface of a non-conducting liquid. For the case of isotropic surface perturbations, the EHD wave turbulence spectrum reads

This work is supported by Russian Science Foundation, project No. 21-71-00006. Corresponding author: E.A. Kochurin.

Igor A. Dmitriev is with Institute of Electrophysics, Ural Branch of Russian Academy of Sciences, 620016 Yekaterinburg, Russia

Evgeny A. Kochurin is with Institute of Electrophysics, Ural Branch of Russian Academy of Sciences, 620016 Yekaterinburg, Russia and Skolkovo Institute of Science and Technology, Moscow, 121205 Russia (e-mail: kochurin@iep.uran.ru)

Nikolay M. Zubarev is with Institute of Electrophysics, Ural Branch of Russian Academy of Sciences, 620016 Yekaterinburg, Russia and P.N. Lebedev Physical Institute, Russian Academy of Sciences, 119991 Moscow, Russia

$$S(k) = C_E \cdot P^{1/2} v_A^{-3/2} k^{-3}, \quad v_A = \left(\frac{\gamma}{\rho}\right)^{1/2} E_0, \quad (2)$$

where $C_E$ is the corresponding KZ constant, $E_0$ is the value of the external horizontal electric field strength, $\gamma(\varepsilon) = \varepsilon_0 (\varepsilon-1)^2 / (\varepsilon+1)$ is an auxiliary coefficient, $\varepsilon_0$ is the electric constant, and $\varepsilon$ is the relative dielectric constant of the liquid. The quantity $v_A$ has a sense of the velocity of waves propagating along the direction of the external field (analogue of Alfvén speed). The spectrum (2) is applicable for the case of strong external field, where the influence of capillary and gravity effects can be neglected. The corresponding energy turbulence spectrum has from:

$$E(k) \approx v_A^2 k S(k) = C_E \cdot P^{1/2} v_A^{1/2} k^{-2}. \quad (3)$$

The spectrum (3) is the direct analogue of the weak magnetohydridynamic turbulence spectra [20-22] for free-surface EHD turbulence.

The evidence of realization of (2) was recently discovered numerically in [23] for MHD wave turbulence on the surface of ferrofluid in anisotropic regime of motion, where dominant direction of wave propagation was perpendicular to the external magnetic field, i.e., $k \sim k_y$. The possible reason for such anisotropy was choice of the resonant pumping which was dependent on the direction of wave propagation. The aim of current work is direct numerical simulating the development of EHD surface wave turbulence under the action of isotropic external forcing. It will be shown for the first time that a system of interacting nonlinear surface waves can pass from the state of dispersive capillary turbulence described by (1) to the regime of quasi-isotropic non-dispersive EHD turbulence described by the spectrum (2) with very high accuracy.

## II. THE MODEL EQUATIONS

The work considers the potential flow of an ideal incompressible non-conducting fluid of infinite depth with a free surface in a uniform external electric field directed tangentially to the unperturbed boundary. It is assumed that there is a non-conducting light gas or vacuum above the fluid surface. We introduce a Cartesian coordinate system with the radius vector, $\mathbf{r} = \{x, y, z\}$. The equality $z = 0$ describes the unperturbed fluid boundary, and the equation $z = \eta(x,y,t)$ determines the profile of the liquid surface. The external electric field is assumed to be directed along the $x$-axis and be equal to $E_0$ in absolute value. The fluid velocity potential $\Phi(\mathbf{r})$ satisfies the Laplace equation $\Delta \Phi = 0$ in the region $z < \eta$. We consider the case of a dielectric liquid (there are no free charges in the liquid). Electric field strength $\mathbf{E}_{1,2}(\mathbf{r})$ is described by electric field potentials $\mathbf{E}_{1,2} = -\nabla \varphi_{1,2}$, where indices "1" and "2" correspond to areas inside the liquid, $z < \eta$, and above its free boundary, $z > \eta$, respectively. The potentials of the electric field satisfy the Laplace equations, $\Delta \varphi_{1,2} = 0$. The boundary conditions for the Maxwell equations in terms of the electric field potentials are written at the free surface $z = \eta(x,y,t)$ as follows: $\varphi_1 = \varphi_2$, and $\varepsilon \partial_n \varphi_1 = \partial_n \varphi_2$ where $\partial_n$ is a derivative along the normal to the free surface. At a distance from the liquid boundary $z \to \mp\infty$, the electric field becomes uniform $\varphi_{1,2} = -E_0 x$. The evolution of the system is described by kinematic and dynamic boundary conditions at $z = \eta(x,y,t)$:

$$\eta_t = \Phi_z - \nabla_\perp \eta \cdot \nabla_\perp \Phi,$$

$$\Phi_t + \frac{(\nabla \Phi)^2}{2} + g\eta - \sigma \nabla_\perp \cdot \frac{\nabla_\perp \eta}{\sqrt{1+(\nabla_\perp \eta)^2}}$$
$$= \frac{\varepsilon_0(\varepsilon-1)}{2\rho}\left(\nabla \varphi_1 \cdot \nabla \varphi_2 - E_0^2\right),$$

where $g$ is the acceleration of gravity, $\nabla$ and $\nabla_\perp$ are the differential operators defined as $\nabla = \{\partial_x, \partial_y, \partial_z\}$ and $\nabla_\perp = \{\partial_x, \partial_y\}$, resepcively. The written equations represent a closed system of equations describing fully nonlinear dynamics of the free surface of a dielectric liquid in an external horizontal electric field taking into account gravity and surface tension. The total energy of the system (Hamiltonian) is expressed as

$$\mathcal{H} = \frac{1}{2}\int_{z\leq\eta}(\nabla\Phi)^2 d\mathbf{r} - \frac{\varepsilon_0\varepsilon}{2}\int_{z\leq\eta}\left((\nabla\varphi_1)^2 - E_0^2\right)d\mathbf{r} \quad (4)$$
$$-\frac{\varepsilon_0}{2}\int_{z\geq\eta}\left((\nabla\varphi_2)^2 - E_0^2\right)d\mathbf{r} + \int\left[\frac{g\eta^2}{2} + \frac{\sigma}{\rho}\left(\sqrt{1+\nabla_\perp\eta}-1\right)\right]dxdy.$$

The dispersion law for linear waves at the boundary of a dielectric liquid in an external horizontal electric field has the form [24]:

$$\omega^2(\mathbf{k}) = gk + v_A^2 k_x^2 + \frac{\sigma}{\rho}k^3. \quad (5)$$

In the absence of an external field, the dispersion relation (5) describes the propagation of surface gravity-capillary waves. Their minimum phase velocity is reached at the wavelength $\lambda_0 = 2\pi(\sigma/g\rho)^{1/2}$ with wave period $t_0 = 2\pi(\sigma/g^3\rho)^{1/4}$. It is convenient to use these quantities as characteristic values of length and time, and introduce dimensionless units as $\tilde{t} = t/t_0$ and $\tilde{\mathbf{r}} = \mathbf{r}/\lambda_0$ (further tilde signs are omitted for brevity). In dimensionless form, the dispersion relation (5) can be rewritten as

$$\omega^2(\mathbf{k}) = k + V_A^2 k_x^2 + k^3,$$

where $V_A^2 = \gamma E_0^2 / (\sigma g \rho)^{1/2}$ is the non-dimensional parameter defining the external electric field strength (non-dimensional Alfvén speed). Further in the work, the first term on the right side of the dispersion law is neglected, which corresponds to the consideration of short wavelengths, $k \gg 1$.

In a weakly nonlinear approximation, the system of electrohydrodynamic equations can be reduced to a pair of equations describing directly the dynamics of the boundary. The procedure for deriving such equations is described in detail in [25-27]. The equations of motion of the liquid surface can be obtained by variational differentiation of the Hamiltonian:

$$\frac{\partial \eta}{\partial t} = \frac{\delta \mathcal{H}}{\delta \psi}, \qquad \frac{\partial \psi}{\partial t} = -\frac{\delta \mathcal{H}}{\delta \eta}, \qquad (6)$$

where the quantities $\eta(x,y,t)$ and $\psi(x,y,t) = \Phi(x,y,z=\eta,t)$ play the role of canonical variables. Taking into account the cubically nonlinear terms in the integrand, the Hamiltonian of the system (4) acquires the form

$$\mathcal{H} = \mathcal{H}_1 + \mathcal{H}_2 + \mathcal{H}_3, \qquad (7)$$

where

$$\mathcal{H}_1 = \frac{1}{2} \iint \left( \psi \hat{k} \psi + V_A^2 \eta_x \hat{k}^{-1} \eta_x \right) dxdy,$$

$$\mathcal{H}_2 = \frac{1}{2} \iint (\nabla_\perp \eta)^2 dxdy,$$

$$\mathcal{H}_3 = \frac{1}{2} \iint \eta \left( (\hat{k}\psi)^2 - (\nabla_\perp \psi)^2 \right) - V_A^2 A_E \left( \eta \eta_x^2 - \eta_x \hat{k}^{-1} \eta \hat{k} \eta_x \right.$$
$$\left. + \eta_x \hat{k}^{-1} (\nabla_\perp \eta \cdot \nabla_\perp \hat{k}^{-1} \eta_x) \right) dxdy.$$

Here $A_E = (\varepsilon - 1)/(\varepsilon + 1)$ is an analogue of the Atwood number for the electric field, $\hat{k}$ is an integral operator defined in Fourier space as $\hat{k} f_\mathbf{k} = k f_\mathbf{k}$, and $\hat{k}^{-1}$ is the operator inverse to $\hat{k}$. The first term $\mathcal{H}_1$ in (7) has a sense of sum of the kinetic and potential electric energy, $\mathcal{H}_2$ determines the energy of dispersive capillary waves, and $\mathcal{H}_3$ corresponds to nonlinear wave interaction energy.

For a complete description of the developed EHD wave turbulence, the terms responsible for the external force (energy pumping) and energy dissipation (viscosity) should be added to the equations (6). As a result, the equation system for the boundary motion takes the following form

$$\eta_t = \hat{k}\psi - \hat{k}(\eta \hat{k}\psi) - \nabla_\perp(\eta \nabla_\perp \psi) + \hat{D}_k \eta, \qquad (8)$$

$$\psi_t = \nabla_\perp^2 \eta + \frac{1}{2}\left[(\hat{k}\psi)^2 - (\nabla_\perp \psi)^2\right] + V_A^2 \hat{k}^{-1} \eta_{xx}$$
$$- \frac{A_E V_A^2}{2}\left[ 2\hat{k}^{-1}\partial_x \left(\eta \hat{k}\eta_x - \nabla_\perp \eta \cdot \nabla_\perp \hat{k}^{-1} \eta_x \right) - \eta_x^2 \right.$$
$$\left. - 2\eta \eta_{xx} - (\nabla_\perp \hat{k}^{-1} \eta_x)^2 \right] + \mathcal{F}(k,t) + \hat{D}_k \psi, \qquad (9)$$

where $\hat{D}_k$ is an operator responsible for the influence of viscosity given in $k$-space as $\hat{D}_k f_k = -\nu(k - k_d)^2 f_k$ for $k \geq k_d$ and $\hat{D}_k = 0$ for $k < k_d$; the coefficient $\nu$ determines the intensity of energy dissipation. The term responsible for the energy pumping of the system $\mathcal{F}(k,t)$ is given in Fourier space as

$$\mathcal{F}(k,t) = F(k)\exp[iR(\mathbf{k},t)],$$

$$F(k) = F_0 \exp[-(k - k_0)^4 / k_f],$$

where $R(\mathbf{k},t)$ are the random numbers uniformly distributed in the interval [0, 2π], and the coefficient $F_0$ determines the maximum pumping amplitude, which is achieved at the wavelength $k = k_0$. The surface perturbations are excited in the following range of wave numbers: $1 \leq k \leq k_f$. In the limit of a strong field and in the absence of dissipation and pumping of energy, there were found [25,26] exact analytical solutions of the equation system (8) and (9) in the form of nonlinear stationary surface waves of an arbitrary shape propagating along the direction of the electric field, i.e., similar to Alfvén waves in a perfectly conducting liquid or plasma. These solutions are applicable for a dielectric liquid with high dielectric constant. At finite $\varepsilon$, the surface waves can collapse under the action of a strong horizontal field [28]. The wave collapse time is minimal for a liquid with $\varepsilon = 5$, see [28] for more details. Thus, the nonlinear effects for liquids with $\varepsilon \approx 5$ are most pronounced. For this reason, all simulations in this work are performed for liquids with this value of dielectric constant. The characteristic electric fields lie in the interval, 1-10 kV/cm [28].

The computational model used in this work is based on the numerical solution of the system of equations (8) and (9). The calculation of spatial derivatives and integral operators is carried out by pseudo-spectral methods with the total number of Fourier harmonics $N \times N$. The algorithm is parallelized on GPU using NVidia CUDA platform. Numerical integration in time is based on the fourth-order explicit Runge-Kutta method with step $dt$. The simulations are performed in a periodic region of size $2\pi \times 2\pi$ with the following parameters: $N = 1024$, $dt = 2.5 \cdot 10^{-5}$, $F_0 = 1.5 \cdot 10^6/N^2$, $k_0 = 3$, $k_f = 6$, $k_d = 150$, $\nu = 10$. To suppress the aliasing effect, we used a filter that nulls higher harmonics with a wavenumber above $N/3$. The spectral methods used for calculating spatial derivatives and integral operators are very accurate. The relative error in determining the energy is estimated in the order of $10^{-12}$, see [29].

## III. SIMULATION RESULTS

This work presents six series of calculations of the nonlinear dynamics of the free surface of a liquid for a wide range of the external electric field strengths: $V_A = 0, 5, 10, 20, 30$, and $40$. These field values were chosen to demonstrate the transition from capillary wave turbulence to the regime of EHD wave turbulence in a strong field. Fig. 1 shows the evolution of the total energy of the system (7) for the different values of $V_A$. It can be seen that the system quickly passes to a quasi-stationary regime of motion, in which the influence of the energy pumping is compensated by dissipation, and the total energy of the system oscillates near some constant value. It can be seen that average energy level does not depend on the value of the field strength. The explanation for this fact is that the system under study is not conservative. With increase the external field the speed of wave propagation increases too leading to more intensive energy dissipation per unite time. The inset to Fig. 1 shows the time dependence of the electric and capillary energies for the maximum value of the field parameter, $V_A = 40$. It can be seen that in the quasi-stationary regime of motion, the total electric energy is much larger than the energy of capillary waves, i.e., the system indeed reaches the regime of developed EHD wave turbulence.

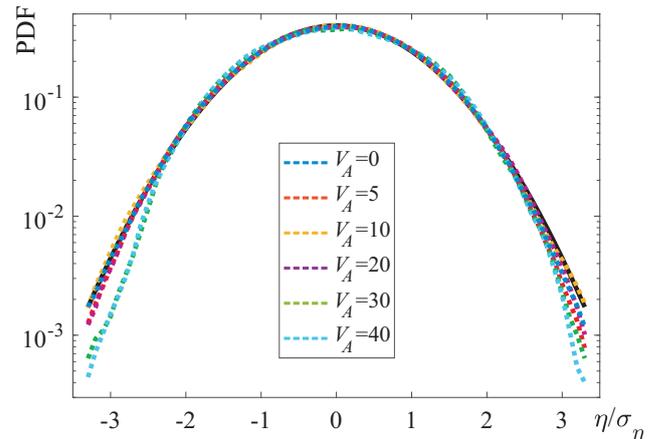

**Fig. 2.** The probability density functions (PDF) for surface elevations rescaled by standard deviation $\sigma_\eta$ are shown at the stationary state for different values of $V_A$, black solid line corresponds to Gaussian distribution.

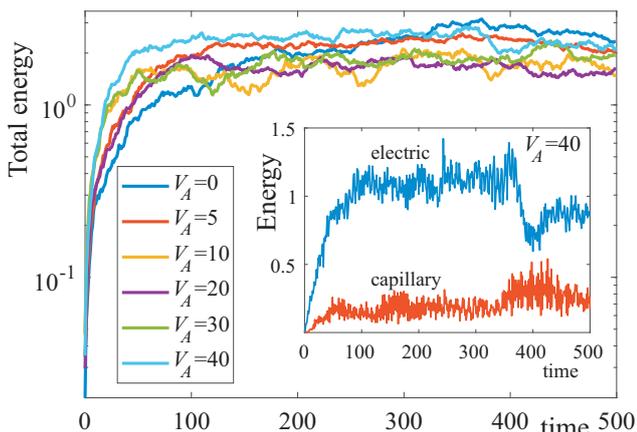

**Fig. 1.** The total energy of the system (7) versus time is shown for the different external fields; the inset shows temporal evolution of the electric and capillary energies for the maximum external field, $V_A = 40$.

Fig. 2 shows probability density functions of the wave elevation rescaled by its standard deviation $\sigma_\eta$ measured at the steady states. As can be seen, the probability densities become close to the normal distribution (the solid line in Fig. 2 corresponds to the Gaussian function), which indicates the realization of the regime of developed wave turbulence on the liquid surface.

Fig. 3 shows the shape of the liquid surface for two limiting cases $V_A = 0$ and $V_A = 40$ at the time moment $t = 500$. It can be seen that the fluid motion has a complex and chaotic character. For a zero field, the motion is completely isotropic. The situation changes for a strong electric field: surface waves propagate mainly in the direction of the $y$-axis, i.e., perpendicular to the external field.

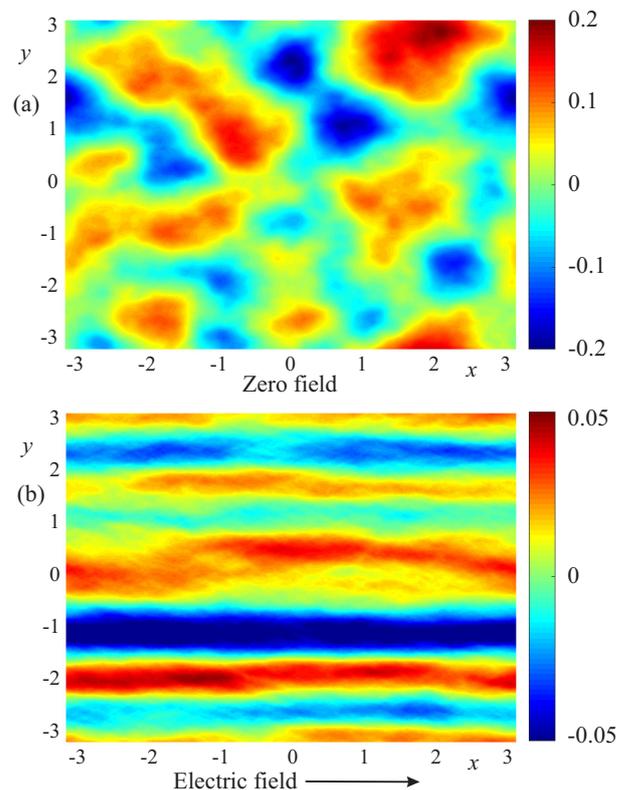

**Fig. 3.** Free surface elevations are shown at the stationary state, $t = 500$, (a) for zero field, $V_A = 0$, and (b) for the maximum external field, $V_A = 40$.

Fig. 4 demonstrates the gradients of surfaces presented in Fig. 3. One can see the same tendency: motion of surface becomes very complicated. In Fig. 4b, we also see structures in the form of packets of capillary waves propagating perpendicular to the external field. The excitation of capillary waves traveling along *y*-axis can be related with the realization of resonant three-wave interactions:

$$\omega(\mathbf{k}) = \omega(\mathbf{k}_1) + \omega(\mathbf{k}_2), \qquad \mathbf{k} = \mathbf{k}_1 + \mathbf{k}_2, \qquad (10)$$

where the dependence $\omega(\mathbf{k})$ is determined from the dispersion relation (5). The relations (10) can be interpreted as conservation laws for energy $\omega$ and momentum $\mathbf{k}$ of interacting waves. Surface perturbations propagating along the direction of the field have a higher energy than capillary waves of the same wavelength traveling along the *y*-axis. The frequency (or energy) of capillary waves increases with the wave number as $k^{3/2}$. Thus, low-frequency EHD waves traveling parallel to the external field can be in resonance with high-frequency capillary waves propagating perpendicular to the field direction. Apparently, the anisotropy observed in Fig. 3 and Fig. 4 arises in result of such resonant wave interactions.

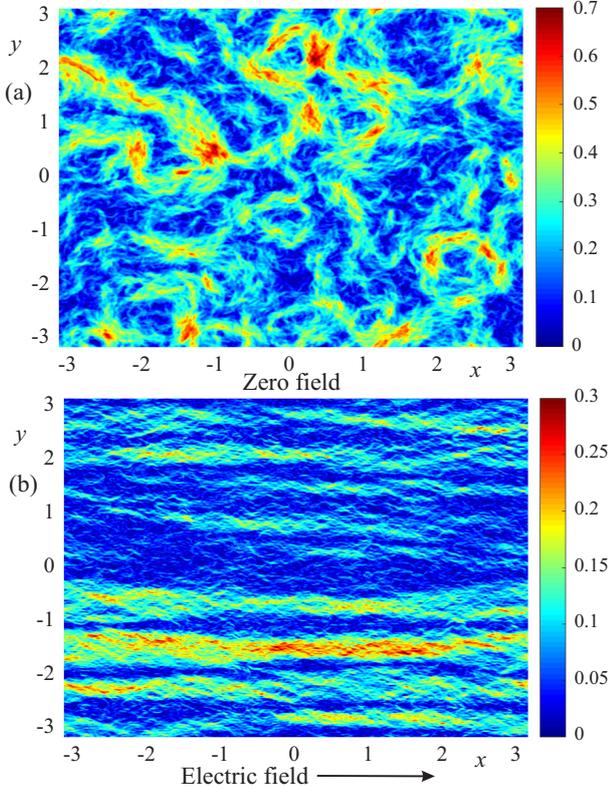

**Fig. 4.** Free surface gradients are shown at the stationary state, $t = 500$, (a) for zero field, $V_A = 0$, and (b) for the maximum external field, $V_A = 40$.

The main question of the current study is whether the obtained results are consistent with the analytical weak turbulence spectra (1) and (2). To answer it, Fig. 5 shows the surface spectra averaged over the phase angle in Fourier space for different values of $V_A$. It can be clearly seen that the presented spectra have a power-law distribution in the inertial interval of wave numbers consisting of more than one decade: $6 \leq k \leq 150$. In the absence of an electric field, the calculated turbulence spectrum agrees with Zakharov-Filonenko's spectrum (1) with very high accuracy. As the strength of the external field increases, we observe a transition to the EHD wave turbulence with the spectrum (2). It can be seen from Fig. 5 that the transition to the spectrum (2) occurs at electric fields of the order $V_A \sim 10$. With a further increase in the field, the slope of the spectrum does not change, and the exponent remains close to $-3$. Thus, the numerical simulation results show that the transition from dispersive capillary wave turbulence to non-dispersive EHD wave regime is indeed possible.

Recall that expression (2) is obtained on the basis of dimensional analysis of the weak turbulence spectra under the assumption of an isotropic character of fluid motion. At the same time, Fig. 3 and Fig. 4 show that the propagation of surface waves in a strong field is anisotropic: waves propagating perpendicular to the electric field dominate. At first glance, we have a paradox: spectrum (2) is obtained for non-dispersive waves, for which, $\omega = k$, but the dominant role is played by the capillary waves with the dispersion law, $\omega = k^{3/2}$. In order to clarify this issue, Fig. 6 shows the energy density of the local electric field at the liquid boundary for the case corresponding to Fig. 3(b) and Fig. 4(b). In Fig. 6, we do not see any preferred directions: the energy is distributed almost isotropically. In fact, the capillary waves propagating along the *y*-axis do not make a significant contribution to the total energy of the system, i.e., $\mathcal{H}_1 \gg \mathcal{H}_2$, which is confirmed in the inset to Fig. 1. Thus, electrohydrodynamic effects completely dominate over the capillary ones.

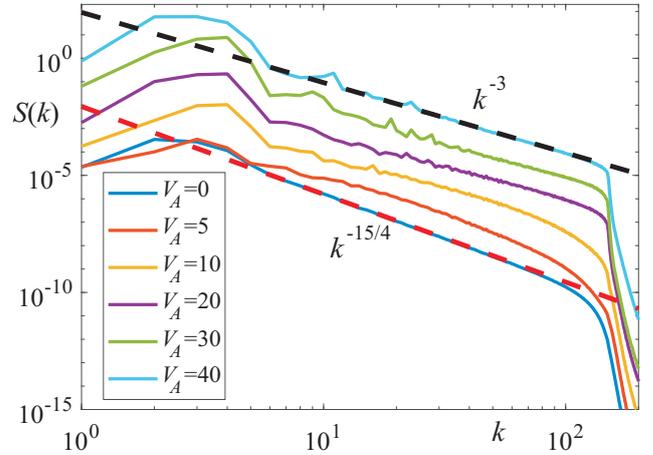

**Fig. 5.** Surface turbulence spectra $S(k)$ are shown for different values of $V_A$ (the dependencies are shifted for clarity). Red dashed line: Zakharov-Filonenko's spectrum (1). Black dashed line: EHD wave turbulence spectrum (2).

In order to track the effects of anisotropy, we plot the Fourier transform of the electric field energy density in Fig. 7. It can be seen that the distribution of energy in *k*-space is almost uniform except for a narrow band along the line $k_x = 0$. The absence of electric energy in this region is explained by its dependency on the shape of surface perturbations:

$$E(\mathbf{k}) = V_A^2 k_x^2 k^{-1} |\eta_{\mathbf{k}}|^2. \qquad (11)$$

As can be seen, the region with $k_x = 0$ is physically distinguished: electric field does not affect waves propagating along the *y*-axis. Thus, the energy turbulence spectrum is formed due to surface perturbations propagating within wide angle along the *x*-axis expect the angles $\pi/2$ and $-\pi/2$, where capillary effects play a dominant role. In conclusion, we note that expression (11) after averaging over the phase angle and passing to polar coordinates in Fourier space gives the energy spectrum (3), i.e., $E(k) \sim k^{-2}$. This dependence differs from the spectrum of two-dimensional weakly dispersive wave turbulence, $E(k) \sim k^{-1}$, recently studied in [30]. The explanation for this difference may lie in the fact that we study a system of almost non-dispersive waves, in contrast to the work [30], where an effect of dispersion is taken into account. We also note that the simulation results do not demonstrate the influence of any coherent structures such as shock waves observed in plane-symmetric geometry [31].

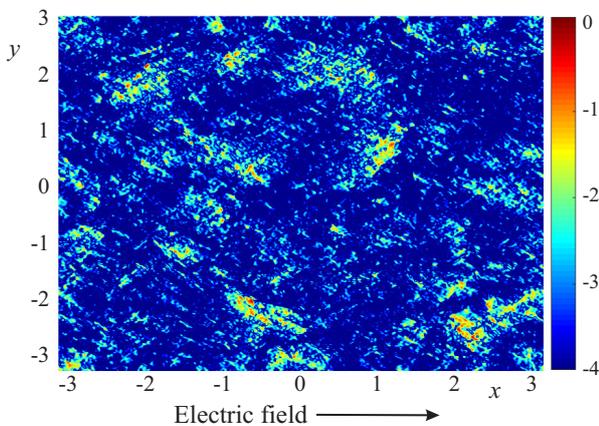

**Fig. 6.** Electric field energy density (logarithmic scale) is shown at the quasi-stationary state ($t = 500$) for the maximum external field, $V_A = 40$.

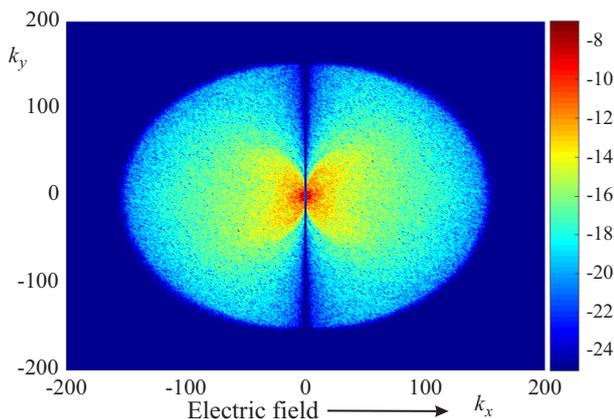

**Fig. 7.** Fourier image of the electric field energy density (11) (logarithmic scale) is shown at the quasi-stationary state ($t = 500$) for the maximum external field, $V_A = 40$.

## V. CONCLUSION

Three-dimensional direct numerical simulations of stochastic motion of dielectric liquid with a free surface under the action of external horizontal electric field are carried out in the current work. The numerical model includes the effects of surface tension, viscosity and external isotropic forcing acting with random phases. The simulations are made for a wide range of the external electric field strengths. The numerical results show that the system of nonlinear interacting surface waves can pass to the stationary chaotic state, which can be interpreted as a developed wave turbulence regime. At the quasi-stationary state, the effects of external forcing are completely compensated by viscosity: the total energy of the system oscillates near some averaged value. At the same time, the fluid motion becomes complex and chaotic. The probability density functions of surface elevation measured in the quasi-stationary state are found to be very close to the normal Gaussian distribution. At the work, a transition from dispersive capillary wave turbulence to a quasi-isotropic EHD surface turbulence with increase of the external electric field strength is observed for the first time. In the regime of developed EHD wave turbulence, the total electrical energy is much higher than the energy of capillary waves, i.e., electrohydrodynamic effects play a dominant role. Anisotropic effects that lead to the generation of capillary waves traveling perpendicular to the external field direction are detected at the regime of developed EHD wave turbulence. Despite this fact, the numerically obtained spectrum of EHD turbulence is in good agreement with the analytical spectrum derived on the basis of the dimensional analysis of weak turbulence spectra. Thus, the simulation results indicate the universal character of weak turbulence theory allowing to describe turbulence spectra in arbitrary nonlinear wave systems. We believe that the results obtained can be useful for developing methods for controlling the movement of fluids and creating a given small-scale relief on their free boundaries.